\documentclass[12pt,preprint,letterpaper]{aastex}

\shorttitle{UV Excess Evolution}
\shortauthors{Atlee, Assef \& Kochanek}

\begin{document}

\title{Evolution of the UV Excess In Early-Type Galaxies}

\author{David W. Atlee, Roberto J. Assef and 
Christopher S. Kochanek\altaffilmark{1}}
\affil{Department of Astronomy, The Ohio State University}
\altaffiltext{1}{Center for Cosmology and Astroparticle Physics,
The Ohio State University}
\email{atlee@astronomy.ohio-state.edu}

\begin{abstract}
We examine the UV emission from luminous early-type galaxies
as a function of redshift.  We perform a stacking analysis using
Galaxy Evolution Explorer (GALEX) images of galaxies 
in the NOAO Deep Wide Field Survey (NDWFS) Bo\"otes field and examine
the evolution in the UV colors of the average galaxy.
Our sample, selected to have minimal
ongoing star formation based on the optical to mid-IR SEDs of the galaxies, 
includes 1843 galaxies spanning the redshift range
$0.05\leq~z\leq0.65$.  We find evidence that the strength of the UV excess 
decreases, on average, with redshift, and our
measurements also show moderate disagreement with previous models of the
UV excess.  Our results show little evolution in the 
shape of the UV continuum with redshift,
consistent either with the binary model for the formation of Extreme Horizontal
Branch (EHB) stars or with no evolution in EHB morphology with
look-back time.  However, the binary formation model predicts that the
strength of the UV excess should also be relatively constant, in
contradiction with our measured results.
Finally, we see no significant influence of a galaxy's environment
on the strength of its UV excess.
\end{abstract}


\section{Introduction}\label{secIntro}
The UV excess is defined as the presence of
more UV flux than predicted for a simple, old stellar population
and was first reported by \citet{code79}.  It is sometimes also
called the UV upturn, because $F_{\lambda}$ is seen to rise
shortward of ${\rm 2500\AA}$ \citep{brow04}.  \citet{dona95}, for example,
found that the average early-type galaxy in the Coma cluster is
more than a magnitude bluer in $m_{UV}-B$ than predicted by the
population synthesis models of \citet{bruz93}.  They explaind the observed
emission by invoking residual star formation (RSF), but it is unlikely
that all of the early-type galaxies in the cluster experienced a recent
burst of star formation at around the same time; this indicates the
need for a source of UV emission not associated with star formation.
Population synthesis models have since suggested a number of 
potential sources for this
emission, usually involving significant mass loss by stars leaving the
main sequence.  The proposed sources include post-red
giant stars, hot horizontal branch stars and post-AGB stars 
(e.g. \citealt{bres94}).
A combination of
population synthesis models and high resolution spectra obtained with the
Far UV Spectroscopic Explorer (FUSE; \citealt{brow02}) and the 
Hopkins Ultraviolet Telescope (HUT; e.g. \citealt{ferg93}) 
suggest that extreme horizontal branch (EHB) stars, also called
hot subdwarfs (sdB), are the objects most likely to give
rise to the UV emission, as the observed spectra closely match 
predictions from a population of EHB stars with various surface temperatures 
\citep{brow04}.

Thus, conventional wisdom indicates that the stars giving rise to
the UV emission are produced by significant mass loss from stellar envelopes
on the red giant branch (RGB).  If EHB stars are formed from stars with
massive winds on the RGB, then the fraction of stars that find themselves
in this unusual stage of stellar evolution, and thus the strength
of the UV excess, should depend on the average metallicity of the 
host galaxy.  Evidence for this picture can be found from several
sources.  First is the direct correlation between the strength of the UV
excess and the Lick ${\rm Mg_{2}}$ spectral index, which measures
a galaxy's average metallicity \citep{burs88}.  
Also in keeping with such a metallicity dependence is
the correlation between the UV excess and the mass (luminosity) of a
galaxy suggested by \citet{ocon99}.
However, this correlation has recently become controversial. 
For example, \citet{rich05}
found no apparent correlation between the ${\rm Mg}_{2}$ index and
the UV excess in a sample of 172 early-type galaxies
from the Sloan Digital Sky Survey (SDSS). 
By contrast, \citet{dona06} found a weak but significant correlation among 
elliptical ($-5.5\leq~T<-3.5$) galaxies, but they found 
no such correlation in lenticular ($-3.5\leq~T<-1.5$) galaxies.
They attribute this difference to the presence of residual star formation in
lenticulars.

Recently, Ree et al. (2007; R07) used Galaxy Evolution
Explorer (GALEX) photometry of galaxy clusters
below redshift $z=0.2$ to measure the evolution in
the UV excess of Brightest Cluster Galaxies (BCGs) of rich clusters
with redshift.  The galaxies
they measured showed no significant evolution, but by expanding their galaxy 
sample with objects studied using HST by Brown
et al. (2000, 2003), they found that the $FUV-V$ colors 
of early-type galaxies in massive clusters become redder
at higher redshift.  
They compare their expanded sample to two evolutionary models, one favoring
sdB formation in metal-rich populations and the other favoring metal-poor
populations, attempting to determine which model agrees
better with the measured colors.  These models are
developed by picking a pair of populations to bracket the $z=0$ galaxies
and then passively ``evolving'' them backwards in time.
They find that both models agree reasonably well with the measurements,
and marginally favor the metal-poor model.  

Han, Podsiadlowski \& Lynas-Gray (2007; HPL) 
suggested that sdB stars might form primarily via
close binary interactions rather than forming via wind-driven mass 
loss on the RGB.  Stars in close binary systems
may eject much of their hydrogen envelope after they evolve off the main
sequence via angular momentum exchange between the envelope
and the binary companion.  Their model makes several specific
predictions, including that the correlation of the UV excess 
with metallicity should be weak, as should the evolution of 
the strength and shape of the UV excess with redshift.
The HPL model is rather appealing, as it can explain the 
limited strength of the
metallicity correlation found in the SDSS galaxies \citep{rich05} and
the large fraction of Galactic field sdB stars found in binary systems 
compared to those found in Galactic globular clusters \citep{cate07}.

The alternative hypotheses for the formation of sdB stars can be tested 
by examining the evolution of the UV colors of
early-type galaxies with redshift.  In this work we measure the 
evolution of the average early type galaxy by stacking GALEX images of
galaxies in the Bo\"otes field of the NOAO Deep Wide Field Survey (NDWFS). 
The galaxies are selected based on their optical to mid-IR spectral 
energy distribution (SEDs), following \citet{asse07}.  We study
the UV emission from elliptical galaxies out to $z=0.65$, 
where the number of sample
galaxies in each redshift bin begins to diminish and the risk of
AGN contamination increases.
In \S\ref{secTarget} we describe our galaxy selection procedures. 
We describe our stacking algorithm and the
analysis of the resulting images in \S\ref{secGalaxy}, refine our
galaxy selection criteria in \S\ref{secRefine}, and we examine the
evolution of the UV excess in \S\ref{secEvolution}.  Finally,
in \S\ref{secConclusion} we consider the consequences of our measurements
for models of sdB formation.

\section{Target Selection}\label{secTarget}
The Bo\"otes field of the NOAO Deep Wide Field Survey (NDWFS) covers
approximately 9 ${\rm deg}^{2}$ centered at
($14^{\rm h}32^{\rm m}$, $+34^{\circ}17'$).
We used the optical
(NDWFS, \citealt{jann99}; zBo\"otes, \citealt{cool07}), near-IR (NDWFS; 
FLAMEX, \citealt{elst06}) and mid-IR (The IRAC Shallow Survey, 
\citealt{eis04}) photometry for objects in the field.  The AGN and Galaxy
Evolution Survey (AGES) redshift catalog has galaxy spectra complete
to $I=18.5$ and an extended sample with $18.5<I\leq20$ for objects in
the Bo\"otes field (\citealt{kochIP}).  The AGES catalog contains
spectroscopic redshifts for approximately 17,000 objects, and we use these
redshifts for the galaxies in our sample wherever possible.  We rely on
photometric redshifts for objects without spectroscopy, which is the
majority of our sample.

The GALEX satellite is conducting a survey of
the ultraviolet sky in two photometric bands, the Far-UV ($FUV$;
$\lambda_{eff}=1528$\AA, $\Delta\lambda=442$\AA) and Near-UV ($NUV$;
$\lambda_{eff}=2271$\AA, $\Delta\lambda=1060$\AA; \citealt{morr05}).  
We acquired GALEX Release 3 (GR3) images of sixteen GALEX Deep 
Imaging Survey (DIS) fields overlap the Bo\"otes field. 
The names and GR3 exposure times for all sixteen fields, 
in both the $FUV$ and $NUV$ bands, are listed in Table \ref{tblPointings}.
The different pointings vary widely in depth for both the $FUV$ and
$NUV$ bandpasses, with a typical
integration time for the $NUV$ fields of approximately 7000s.
The $FUV$ pointings are fewer and their exposure times more widely 
distributed.

\citet{asse07} developed and tested a set of three moderate-resolution
template galaxy spectra extending from 0.2 to 10$\mu{\rm m}$.
We used these templates to model the SEDs of the galaxies in the Bo\"otes
field with $I<21.5$ mag by fitting the measured optical, NIR and MIR
fluxes.  The templates are able to distinguish between passively evolving
galaxies and galaxies with on-going star formation based on their MIR fluxes,
which are extremely sensitive to the PAH emission features associated
with star formation.
The MIR fluxes are also sensitive to the presence of AGN since the power-law
continuum typical of AGN results in much stronger $3.6-4.5\mu {\rm m}$
emission than is typical for a stellar population \citep{ster07}.

Using the template spectra, we identified
passively evolving galaxies, eliminated AGN, computed K-corrections and 
synthesized unmeasured bands.  We eliminated AGN from our galaxy
sample by first removing any object flagged as
an AGN or candidate AGN by AGES and accepting the remaining galaxies only if
their photometry fit the galaxy templates with $\chi^{2}_{\nu}\leq2.0$.
AGES targeted galaxies as AGN candidates if they were identified as
X-ray sources in the xBo\"otes survey \citep{murr05}, as radio sources
by FIRST \citep{beck95} or by \citet{devr02}, or as possessing a 
red mid-IR color \citep{ster07}.
The AGES database further identifies galaxies as AGN by the spectral
template used for estimating redshifts in a modified version of
the Sloan Digital Sky Survey spectroscopic pipeline.
Our $\chi^{2}_{\nu}$ cut would eliminate the objects
flagged based on their optical and MIR colors, but it is also able to
identify AGN based solely on how well the measured fluxes agree with a
purely stellar origin across a wide wavelength baseline (i.e. whether the
$B_{\rm w}$ and IRAC 8$\mu {\rm m}$ fluxes can simultaneously agree 
with the $I$-band fluxes).

\citet{asse07} classified
galaxies by the fraction of their bolometric luminosity contributed
by the elliptical component of their SED, designated $\hat{e}$.  We selected an
initial sample of early-type galaxies by requiring
$\hat{e}\geq0.8$, which \citet{asse07} found
roughly divides the Red Sequence the Blue Clump for
galaxies in the Bo\"otes field.  
We also required that galaxies in our sample have ``bolometric'' luminosities,
computed using the model SEDs, in the range 
$0.5\leq\log\biggl(\frac{L_{\rm bol}}{10^{10}L_{\odot}}\biggr)\leq1.5$.
This roughly corresponds to $-24\leq {\rm M}_{R}\leq-21.5$.
After applying these selection criteria, we eliminated an additional seven
galaxies whose MIR fluxes exceed their $K_{s}$ fluxes, since
these galaxies may host hidden AGN or are otherwise unusual.  These
criteria yield an initial
sample of 6630 galaxies in the range $0\leq z\leq0.65$.

\section{Galaxy Stacking}\label{secGalaxy}
We divided our galaxy sample into redshift
bins of width $\Delta z=0.1$, using spectroscopic
redshifts from AGES where available and photometric redshifts from fits
to the \citet{asse07} spectral templates otherwise (approximately 80\% 
of the initial sample).  The photo-z algorithm we employ has been extensively
discussed in \citet{asse07}. For early-type galaxies,
it is accurate to $\Delta z/z\approx0.02$ based on comparisons
to AGES spectroscopic redshifts (see their Fig. 9).

Of the 6630 galaxies with $\hat{e}>0.80$, 328 (122) were detected 
as individual GALEX sources in the $NUV (FUV)$ and
appear in the GALEX catalogs; the majority of these belong to the
first two redshift bins.  Since the vast majority of galaxies in
the sample had no measured UV fluxes, we employed a stacking analysis
to measure the exposure-weighted mean UV fluxes of our galaxy sample as
a function of redshift.  One obvious disadvantage of this approach
is that we are insensitive to variations in individual galaxy properties.
For example, we will be unable to measure the 
presence or strength of any correlation between the UV excess 
and metallicity.

We stack the GALEX images of the Bo\"otes field, 
as described in \S\ref{secImage},
to measure the average UV fluxes of our sample galaxies.  We also
require optical fluxes to compare with the UV emission.
We obtain these by computing exposure-weighted averages of the measured
optical, NIR and MIR fluxes for the galaxies in our sample. 
Since the Bo\"otes photometry is much deeper than the GALEX pointings, the
statistical errors on the appropriately averaged optical fluxes are much
smaller than the uncertainty on the stacked GALEX magnitudes; in fact,
the systematic uncertainties associated with the Bo\"otes photometry dominate
the error budget of the optical, NIR and MIR photometry.
In order to account for the systematic uncertainties, we assign the
averaged magnitudes an uncertainty of 0.05 magnitudes 
before fitting to the spectral templates (see \S\ref{secPhot})
unless the statistical uncertainty implied by averaging the fluxes
exceeds this value.  The statistical uncertainties only exceeded this
systematic limit in the case of the $K_{s}$ band observations, for which
a small number of galaxies with 
exceptionally large uncertainties dominate the error budget.

\subsection{GALEX Image Stacking}\label{secImage}
We obtained the GALEX observations
for each DIS field as well as the associated source catalogs
from the GALEX archive at the Space Telescope Science 
Institute\footnote{http://galex.stsci.edu/GR2/?page=tilelist\&survey=dis}.
The pixel scale of these images is $1\farcs{5}$ per pixel.  We use the
{\it Funtools}\footnote{http://hea-www.harvard.edu/saord/funtools/}
package to parse the images and manage the stacking.
We masked identified GALEX sources falling outside an annulus with
diameter equal to twice the FWHM of the PSF and
centered on the nearest target galaxy;
this guaranteed that the stacked images had a well-defined sky flux.  It also
means that we masked differently in the $FUV$ and $NUV$ images,
both because the PSF size differs between the two bands (4\farcs{5} and 
6\farcs{0} in the $FUV$ and $NUV$, respectively) and because there are
more sources in the $NUV$.
We extracted a list of identified sources from the GALEX
catalog for comparison with the sample
galaxies belonging to each redshift bin.  We identified a set of 
GALEX objects to be masked by comparing the central coordinates
of the objects in the two lists and created separate masked images for
each redshift bin.

Once we identified a GALEX source to be masked, we examined a series of
square frames expanding outward from the center listed in the catalog,
determining whether each individual pixel in the frame needed to be masked.  
We masked pixels whose fluxes exceeded the sky 
background by more than $1\sigma$ and all those within the FWHM of
the PSF.  If more than half of the pixels in a given frame were
masked, we expanded the masking region by one pixel in each direction
and processed the next frame.  In all cases we terminated the 
process if the masking region reached 50 pixels from the center.  The
flux in masked pixels was set to the nominal sky flux in the
appropriate band \citep{morr05}.
We deliberately allowed the masking of flagged objects to 
extend over nearby objects, as this limited the contamination of the
stacked images by stray flux from nearby objects.  It also
removed some flux from the target galaxies, which can introduce a
bias.  We tested the algorithm and found the effects of this bias
to be small (\S\ref{secTest}).
A visual comparison of the GALEX images before and after masking indicated
that the masking algorithm was quite efficient.  

\begin{figure}
\epsscale{1.0}
\plotone{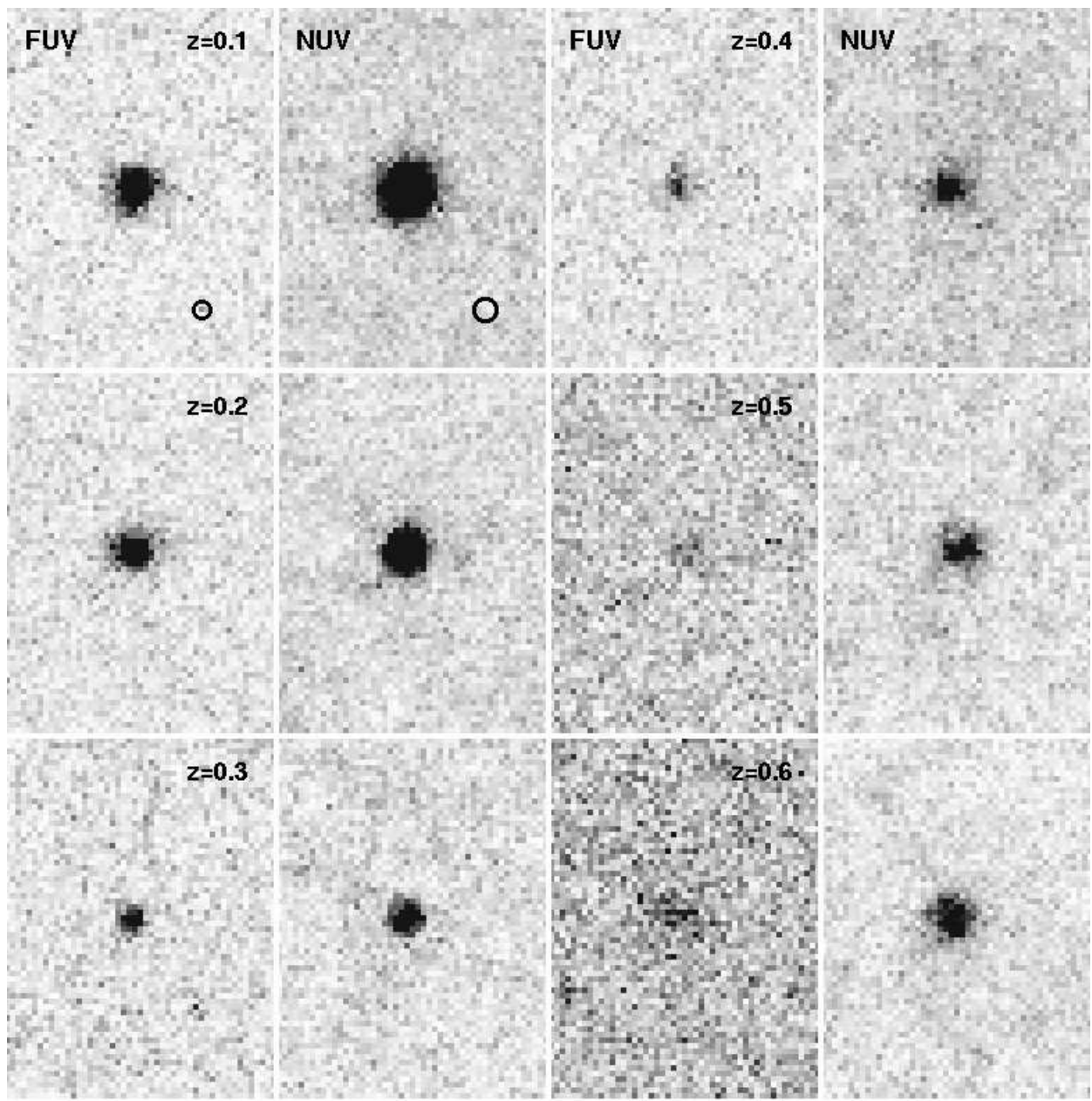}
\caption{Stacked galaxy images.  The first and third columns show the
$FUV$ images while second and fourth show the $NUV$ images.  The redshift 
increases first down columns and then across rows, as indicated.  Each
frame shows a region at the center of the stacked image that is approximately
$45\times65$ pixels, where the plate scale is 1\farcs{5}
per pixel.  The $FUV$ PSF FWHM of 4\farcs{5} and the $NUV$ PSF
FWHM of 6\farcs{0} are indicated by the circles in the $z=0.1$ images.
The images in the higher redshift bins appear to be extended due to jitter
in the positions of the individual sample galaxies caused by finite pixel
sizes.
\label{figStacked}}
\end{figure}

After converting the masked images from counts ${\rm s}^{-1}$ to
counts, we added the
counts in an $81\times81$ pixel ($121\farcs{5}\times121\farcs{5}$)
box around each galaxy in a given redshift bin.  We divided the
number of counts in each stacked pixel by the total exposure time,
converting the
counts back to counts ${\rm s}^{-1}$.  The masking and stacking 
procedures were repeated for each of six redshift bins, centered
from $z=0.1$ to $z=0.6$.  Our stacked galaxy images from each redshift bin
are shown in Figure \ref{figStacked}.  All of the stacked images beyond
the $z=0.1$ redshift bin are consistent with the contributing objects
being unresolved, as expected for the angular resolution of GALEX.

The choices we made in setting the parameters of our masking algorithm
were very conservative, especially the decision to mask pixels down to
$1\sigma$ above the mean sky level.  These choices
were necessitated by the very large number of galaxies, often several
hundred, that go into a single stacked image.  In order to insure that
we measure the sky flux correctly and that we do not introduce stray flux from
the outskirts of nearby sources into the stacked galaxies, we require
our masking algorithm to err on the side of caution.  This decision makes 
it easier to address any intrinsic contamination in our sample.

\subsection{UV Photometry}\label{secPhot}
We performed our UV photometry using the IRAF {\it phot} program with
$15''$ diameter photometric apertures.
We converted the measured count rates to magnitudes using the GALEX
photometric zero points \citep{morr05}:
\begin{equation}\label{eqFuvMag}
m_{FUV} = \log(f_{FUV}) + 18.82
\end{equation}
\begin{equation}
m_{NUV} = \log(f_{NUV}) + 20.08
\end{equation}
where $f_{X}$ is the count rate in band $X$.  We also obtained
optical, NIR and MIR fluxes in the same aperture from the NDWFS,
FLAMEX and IRAC Shallow surveys where available.

Our large photometric aperture is required by the irregularity of the
GALEX PSF, which differs between bands and depends on the
position of a source in the image \citet{mart05}.
Using a Moffat PSF profile with $\beta=3$, which is a reasonable match
to the GALEX PSF, we computed aperture
corrections for our $FUV$ and $NUV$ magnitudes.  Our targets were effectively
point sources beyond $z=0.1$, so a $15''$
aperture includes 99\% and 97\% of the $FUV$ and $NUV$ fluxes, respectively.
These translate to aperture corrections of 0.01 and 0.02 magnitudes
for the $FUV$ and $NUV$ bands, respectively.  Both corrections
are significantly smaller than the errors in the mean stacked magnitudes, 
so we neglect them.

We computed Galactic extinction corrections using the polynomial 
extinction law of
\citet{card89} based on the mean E($B-V$) of 0.011 for objects in the 
Bo\"otes field
\citep{schl98}.  Due to the rapid changes in extinction
across the GALEX photometric bands, we computed a weighted average of the
${\rm R}_{\lambda}$ values across each GALEX bandpass,
\begin{equation}\label{eqExtinction}
R_{\rm x} = \frac{\int^{\lambda_{2}}_{\lambda_{1}} R(\lambda)T(\lambda)d\lambda}
{\int^{\lambda_{2}}_{\lambda_{1}} T(\lambda)d\lambda},
\end{equation}
where $R(\lambda)$ is the \citet{card89} R-value at
wavelength $\lambda$, $T(\lambda)$ is the filter bandpass, and the
extinction in band X is given by
$A_{\rm x}=R_{\rm x}\times E(B-V)$.
Using Eq. (\ref{eqExtinction}), we found ${\rm R}_{FUV}=8.24$ 
and ${\rm R}_{NUV}=8.10$.

We used bootstrap re-sampling of our galaxies to estimate the uncertainty
on the mean magnitude.  This procedure 
naturally includes both counting statistics
and the effects of intrinsic scatter in the sample population, which
could easily cause the uncertainty in the mean magnitude to exceed 
the intrinsic photometric uncertainty.  Any factor that leads to 
variation in $L_{UV}$ at fixed $L_{bol}$ might contribute to the
measured scatter.  Such variables include metallicity \citep{burs88},
age \citep{bres94}, recent star formation \citep{kavi06} and total
stellar mass \citep{ocon99}.  Our bootstrapping 
analysis folds all such intrinsic variations in our galaxy sample into
the error on the mean magnitude.
We drew 250 bootstrapping realizations in each 
redshift bin and used the
RMS of the resulting magnitudes as the uncertainty on the mean magnitude. 
The results of these calculations are listed in 
Table \ref{tabSystematics}.

\begin{figure}
\epsscale{1.0}
\plotone{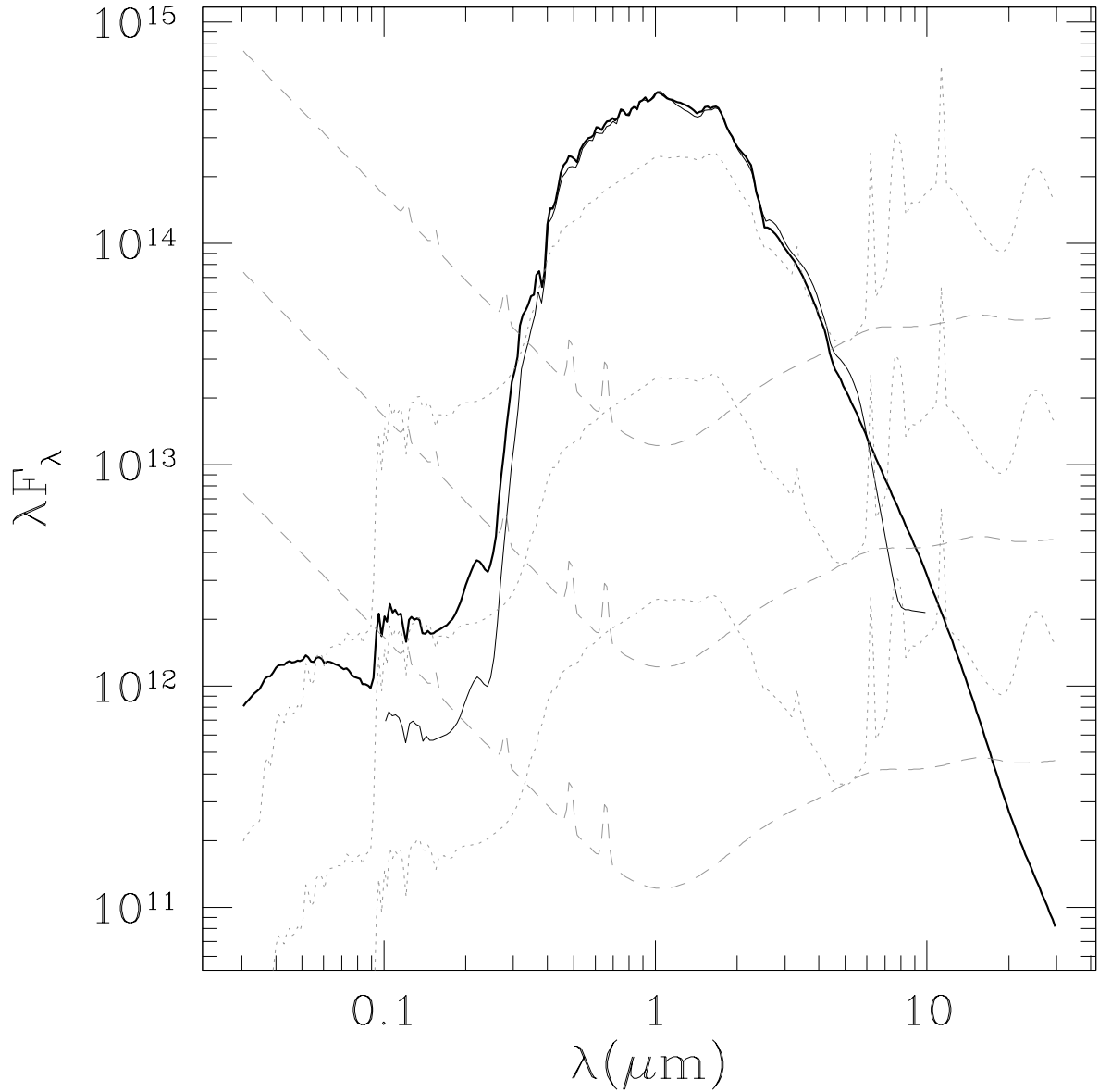}
\caption{The elliptical template we use to compute K-corrections 
({\it thick}), which has been modified from the original ({\it thin}) 
\citet{asse07} elliptical template, in internal template (arbitrary) units.
Also shown are the AGN ({\it dashed}) and Sbc ({\it dotted}) templates,
normalized to contain 1\%, 10\% and 100\% of the bolometric luminosity
contained in the elliptical template.
\label{figTemplate}}
\end{figure}

We did not use the \citet{asse07} spectral templates in our final
analysis because they were computed without using UV photometry to
constrain the shapes of the spectral templates beyond 
$\lambda\approx3000{\rm \AA}$.
Instead, we rely on new, unpublished templates \citep{asse08} that have
been modified by employing GALEX photometry and MIPS $24\mu m$ fluxes
to provide additional constraints on the shapes of the templates.
These templates were developed using the techniques described
in \citet{asse07} and employing the additional UV and MIR photometry.
The new
templates significantly improve the quality of the fit to our stacked UV
data, and we therefore rely on them for UV K-corrections.  The
new elliptical template is compared with the published version in Figure
\ref{figTemplate}.  The figure also shows the spiral and (new) AGN templates
for comparison.  It is apparent that the contributions of AGN and
star-formation can be significant, and it is important that we eliminate
these contaminants wherever possible and subtract the contribution to the
measured UV fluxes where they cannot be eliminated.
(See \S\ref{secEvolution}.)

\subsection{Stacking Tests}\label{secTest}
We conducted two tests to verify that our stacking
code behaved as expected.
First, we tested whether the masking algorithm affected the measured
fluxes from the the stacked galaxies, whether by removing flux from the
target galaxies or by adding extra flux through poor masking of nearby sources.
We divided the 122 galaxies with identified GALEX counterparts in both the $FUV$
and $NUV$ into redshift bins and stacked them.  We compared the magnitudes
of our stacked images with the magnitudes predicted using the
fluxes in the GALEX catalog.
The results, listed in Table \ref{tabSystematics}, suggest
a small bias of approximately 0.05 mag, which is similar to the typical
uncertainty on the bias. 
Also included in Table \ref{tabSystematics}
are the dispersions about the mean magnitude and the estimated bias, showing
that any bias in the measured fluxes is small compared to the
intrinsic scatter in the measured fluxes.

\begin{figure}
\epsscale{0.9}
\plotone{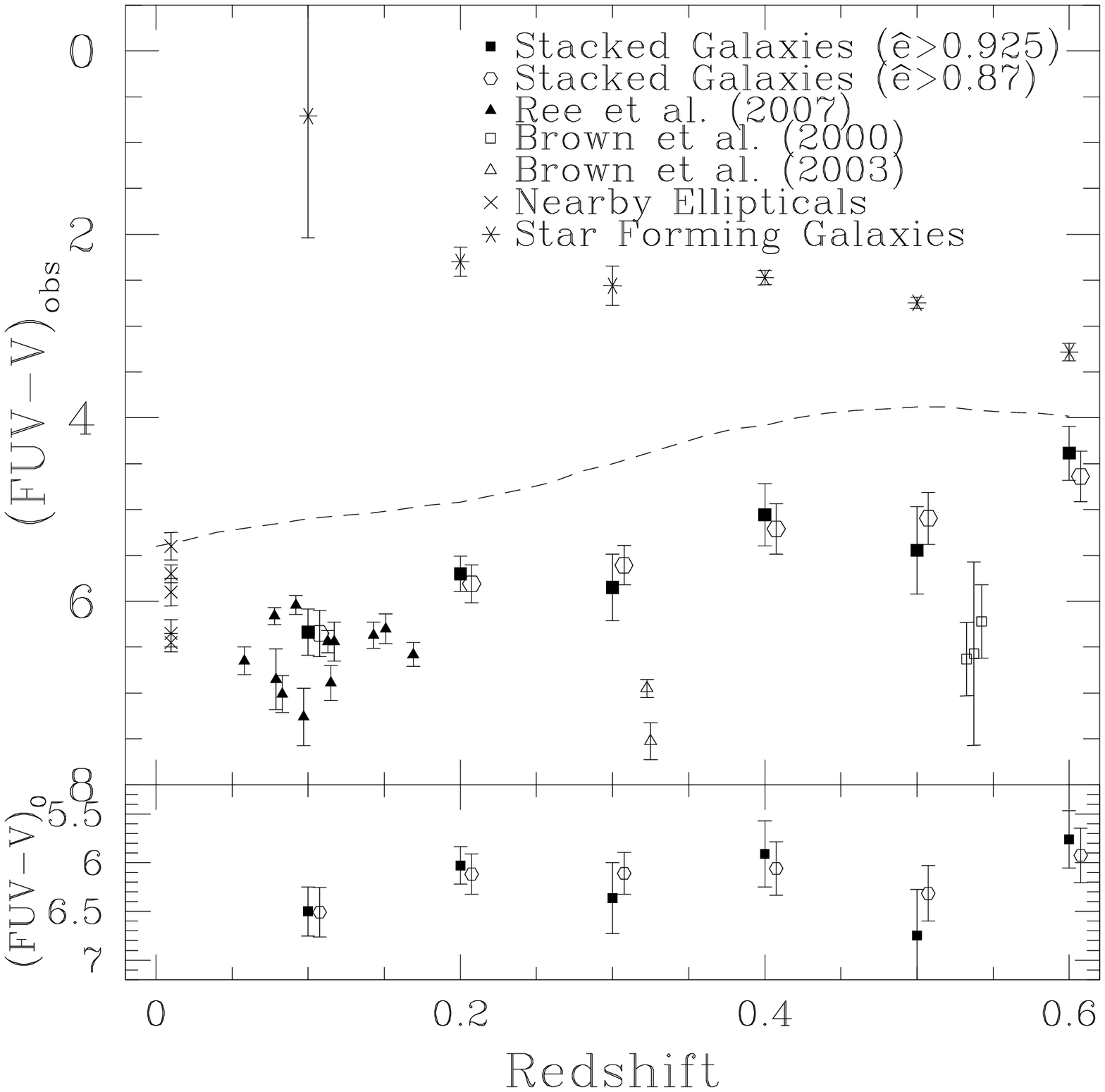}
\caption{Evolution of the observed UV-optical colors
of the stacked early-type galaxies ($\hat{e}\geq0.925$),
compared with the colors of galaxies presented in \citet{ree07}.
{\it Filled squares} and {\it open hexagons}
indicate colors from our stacked galaxies,
{\it open squares} from \citet{brow00},
{\it open triangles} from \citet{brow03},
{\it filled triangles} BCGs from \citet{ree07},
{\it crosses} elliptical galaxies from the Fornax
and Virgo clusters, and {\it stars} stacked star-forming galaxies.
The {\it dashed line} shows the color that would be measured from NGC 1399 
as a function of redshift.  The lower panel shows colors
K-corrected to redshift zero.  The error on the $FUV$ magnitudes
is determined using the dispersion about the average bootstrapped $FUV$
magnitude.  Error bars show $1\sigma$ uncertainties.
\label{figOptColor}}
\end{figure}

We also repeated our entire analysis chain on a set of $\sim8000$
galaxies selected to be strongly star-forming---galaxies
with an elliptical contribution to their bolometric luminosity of
less than 20\%.  We compared the colors of the
stacked star-forming galaxies to those of passively evolving
galaxies, as shown in Figure \ref{figOptColor}.  (Figure \ref{figOptColor}
was created using the photometry of the final galaxy sample, as discussed
in \S\ref{secRefine}, and uses $V$ magnitudes computed using the
template SEDs.)
We compute the uncertainties on our colors using
the bootstrap uncertainties for the $FUV$ magnitudes and setting
$\sigma_{V}=\sqrt{\left[(Bw-Bw_{\rm model})^{2} + (R-R_{\rm model})^{2}\right]/2}$.
As expected, the average star forming
galaxy was significantly bluer than the average elliptical galaxy.
Furthermore, the colors of the stacked galaxies agree well with
the colors of the template spectra at low redshifts and
show different evolution in their $FUV-V$ 
colors compared to the early-type sample. This indicates that our stacking
procedure does not introduce a bias toward bluer $FUV-V$ color 
at high redshift.  This is significant because, as apparent from
Figure \ref{figOptColor}, the $FUV-V$ colors of our stacked galaxies become
somewhat bluer with increasing redshift.

\section{Refining the Galaxy Sample}\label{secRefine}
We selected our initial sample based on the results of
\citet{asse07}, who found that galaxies with $\hat{e}\geq0.80$ fall
on the red sequence.  However, the UV photometry 
we used is more sensitive to low
levels of star formation than optical photometry (e.g. \citealt{kavi06}),
and we needed to determine whether any of our selection criteria
introduce an obvious bias in our galaxy sample.  We therefore broke our
galaxies into subsamples according to various properties and 
looked for any significant differences between them.

\begin{figure}
\epsscale{1.0}
\plotone{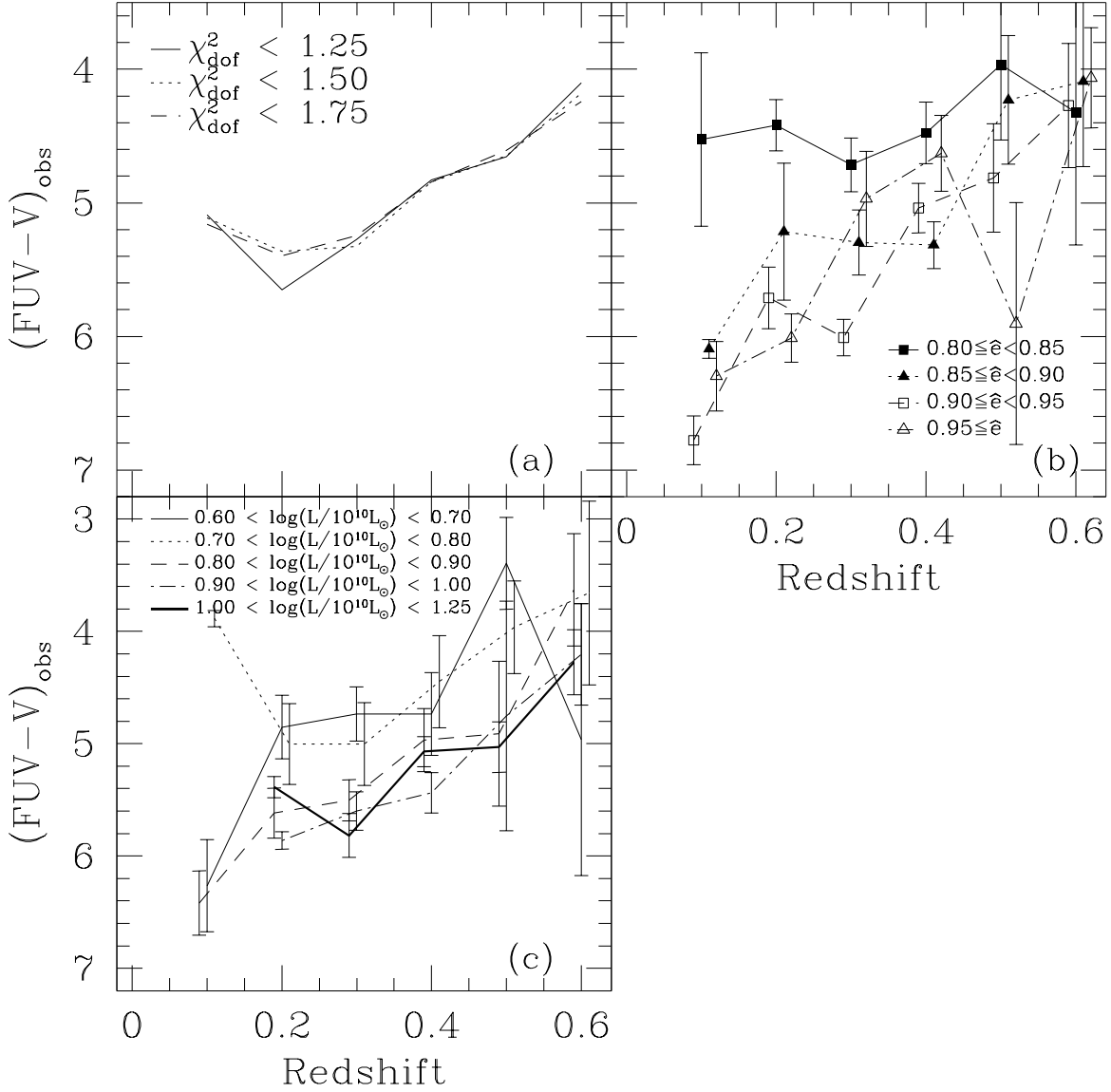}
\caption{Dependence of the redshift evolution of the stacked 
galaxy color on the
(a) goodness of SED fit, (b) elliptical fraction and 
(c) bolometric luminosity.  The bolometric luminosity is 
determined by integrating the galaxy templates.  It is clear
that the trend shown in Fig. \ref{figOptColor} is largely independent
of the selection criteria.  A trend toward redder $FUV-V$ with
higher luminosity is apparent.  Truncated lines in Figs.
(b) and (c) are due to a lack of galaxies in the missing redshift bins.
\label{figSelection}}
\end{figure}

We examined the effect of our three selection criteria---$\hat{e}$, luminosity
and the $\chi^{2}_{\nu}$ of the fit to the template spectra---on 
the redshift evolution of the observed $FUV-V$ color
of the stacked galaxies, which is an indicator for the
strength of the UV excess.  The results are shown in Figure 
\ref{figSelection}.  There is no significant bias associated with 
$\chi^{2}_{\nu}$.  While a small trend with luminosity is observed,
it is only marginally significant.  Also, the sense of the trend is to
introduce a systematic shift toward redder $FUV-V$ at all redshifts.
This does not affect the observed global trend, and is therefore little
cause for concern.  However, it should be noted that a sample with different
$\langle L_{bol}\rangle$ will show slightly different colors.
Examining the variation in $FUV-V$ between samples binned in stellar
mass would not yield additional information because we have
selected only early-type galaxies, so there will
be little variation in mass-to-light ratio.

\begin{figure}
\epsscale{1.0}
\plotone{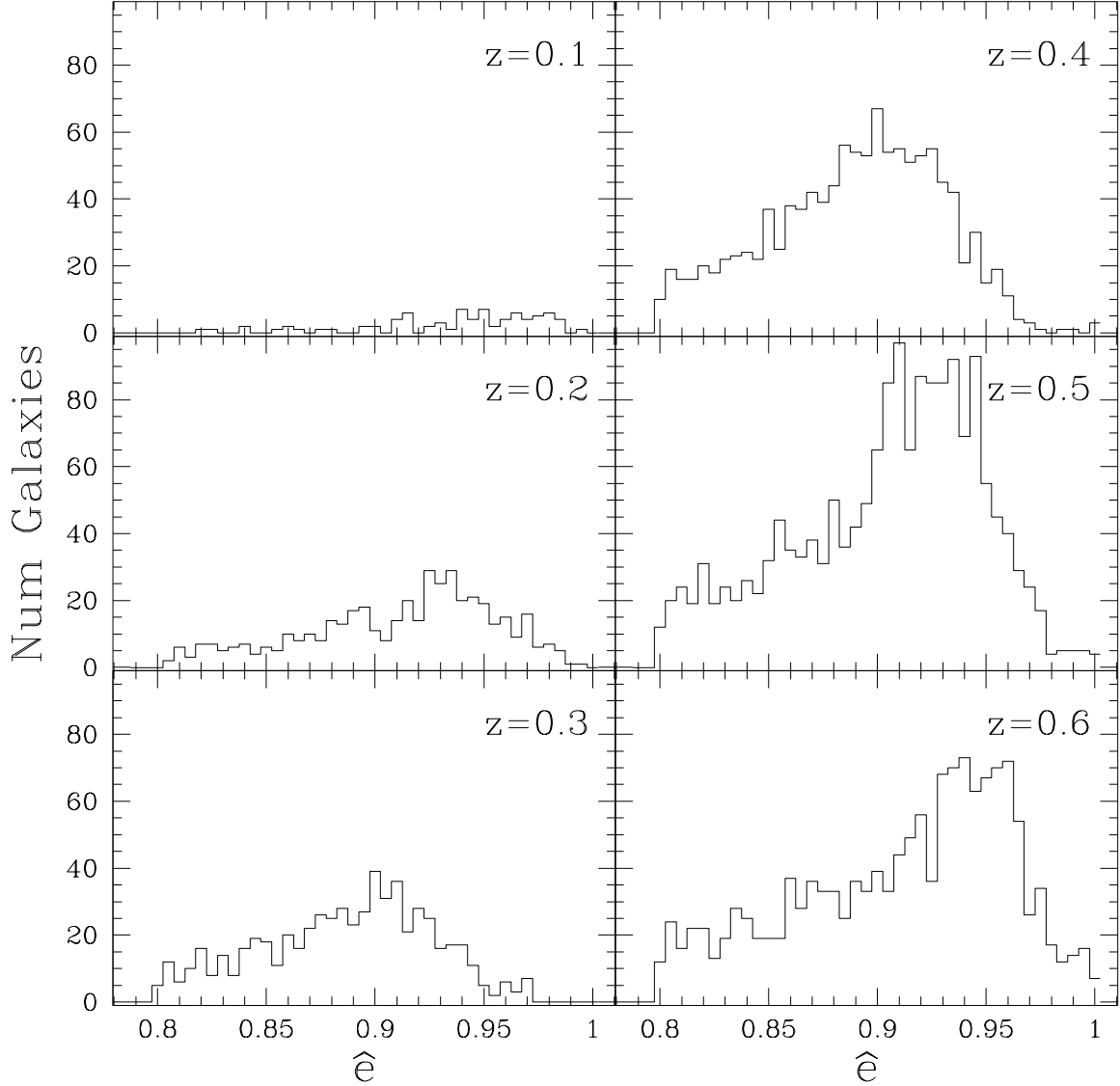}
\caption{The distributions of the $\hat{e}$ parameter in our different
redshift bins for the initial galaxy sample ($\hat{e}>0.80$).  The fraction
of blue (low $\hat{e}$) galaxies increases slowly with redshift,
indicating an increase in fraction of young stars
out to $z=0.5$.  In the $z=0.5$ and $z=0.6$ redshift bins,
the IRAC bands are no longer sensitive to PAH emission from star-forming
galaxies, and we see an increased fraction of sample galaxies at
moderate to high $\hat{e}$.
\label{figEhatDist}}
\end{figure}

The $\hat{e}$ test suggests that the initial $\hat{e}>0.80$ limit 
is too loose.  For the rest of our analysis,
we used a stricter $\hat{e}>0.925$ limit to reduce the
contribution of recent star formation as much as possible.  We
compared the results using this criterion to those
obtained using $\hat{e}>0.87$.  
The new $\hat{e}>0.925$ ($\hat{e}>0.87$) limit left 1843 (4943)
of the original 6330 galaxies.
The uncertainties in our optical and IR photometry are still 
dominated by systematic issues in both cases.  Figure \ref{figEhatDist}
shows the distribution of galaxies in $\hat{e}$ as a function of
redshift.  The distribution broadens toward higher redshift largely
because the redshift of the PAH emission features 
(see Fig. \ref{figTemplate}) reduces our ability to detect low-level
star formation.  Some of the increase in the fraction of
low-$\hat{e}$ galaxies is due to evolution in the stellar
populations of the sample galaxies.  Note that the peak in the 
$\hat{e}$-distribution only moves by $\Delta\hat{e}\approx0.03$ 
between panels.  A strict $\hat{e}$ limit 
reduces the contribution from both of these effects,
but it cannot eliminate them.  As a result, Figure \ref{figTemplates}
shows an increased contribution from the Sab template at higher redshift.  

The magnitudes we measured
from our stacked images are listed in Table \ref{tblPhot} along with
the associated errors and K-corrections.
We examined the evolution of
$\left<\hat{e}\right>$ with redshift and found no significant trend in
either the $\hat{e}\geq0.925$ or $\hat{e}\geq0.87$ cases.  
We verified that the choice between the two alternative selection criteria
has little effect on our conclusions by comparing
the results obtained using the two $\hat{e}$ selection criteria,
which effectively correspond to two different rest-frame color cuts.
We could adopt an $\hat{e}$ limit that varies with redshift
to allow for evolution in the optical 
properties of early-type galaxies, but we choose
not to do so because it would also lead to changes in our sensitivity to
star formation and low-level AGN contamination.  

We also stacked the AGES spectra of the galaxies in bins of
$\hat{e}$ to determine how significantly contamination by
star formation or low-level nuclear activity may contribute to our results.
We divided the sample into three different $\hat{e}$ bins and examined
the spectra in the vicinity of the {\sc [oii]} 3727\AA, 
{\sc [oiii]}4959,5007\AA\ and H$\alpha$ emission lines.
The stacked spectra are shown in Figure \ref{figSpectra} and
indicate the presence of very weak LINERs, which we could not
have detected in a typical individual spectrum.  
The agreement between the mean and median spectra indicates that the 
features are common to most of the galaxies in our sample rather than being
restricted to a few unusual objects.  The strength of the features also
decreases with increasing $\hat{e}$, further motivating the use of the
stricter $\hat{e}>0.925$ limit.  We have now exhausted our ability 
to reduce the presence of weak AGN,
and we must consider this minimal contamination in the 
interpretation of our results.

One additional source of contamination that might be important, especially
at higher redshifts, is blending of the UV light in galaxies
with nearby star-forming companions due to the large scale of the GALEX
PSF.  To assess this possibility, we searched the optical catalogs
for objects within 
4\farcs{5} (6\farcs{0}) of our $\hat{e}\geq0.925$ sample and with 
$I\leq22.5$; beyond these radii, any object with a significant UV
flux would have been masked.  We found 43 (226) total matches to the 
1843 objects in the sample.  If we fit the photometry of the 
companions with the \citet{asse08} templates, compute their
rest-frame colors and require that any ``contaminating'' companions
have $B-R$ at least 0.1 mag bluer than a pure elliptical, 
only 8 (42) companions survive.
We eliminated those galaxies with a blue companion 
and repeated our analysis.  The resulting UV fluxes were
consistent with the main sample given the uncertainties.

\section{Redshift Evolution}\label{secEvolution}
\begin{figure}
\epsscale{1.0}
\plotone{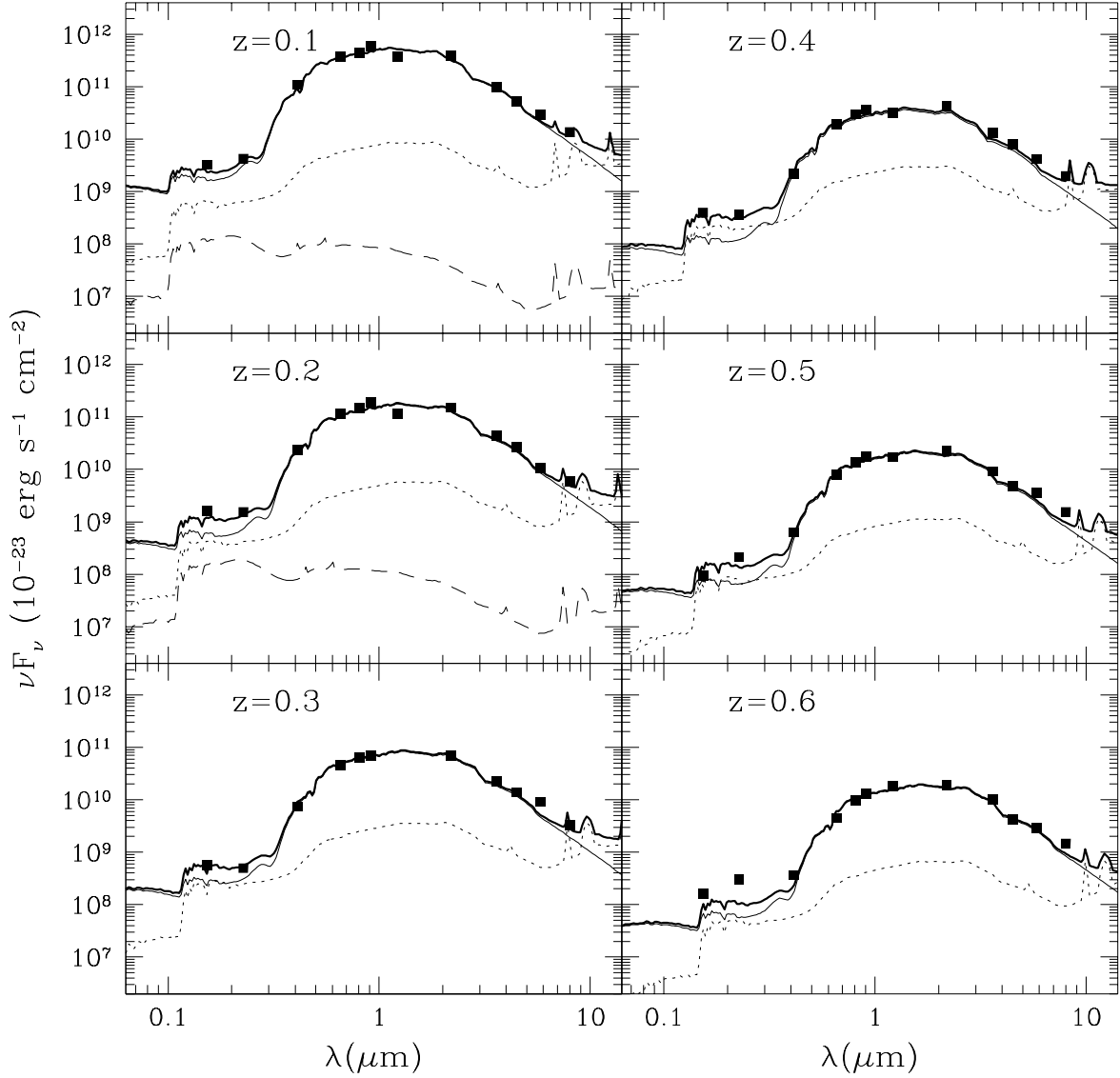}
\caption{Fluxes and model SEDs for stacked galaxies from the
$\hat{e}>0.925$ sample in all redshift bins.  
The {\it dotted} line indicates the spiral component, 
the {\it dashed}
line the irregular component and the {\it thin} line the elliptical
component of the model spectra.  The {\it heavy} line is the sum of
the three components, and it is
used to compute model magnitudes and K-corrections.
\label{figTemplates}}
\end{figure}

\begin{figure}
\epsscale{1.0}
\plotone{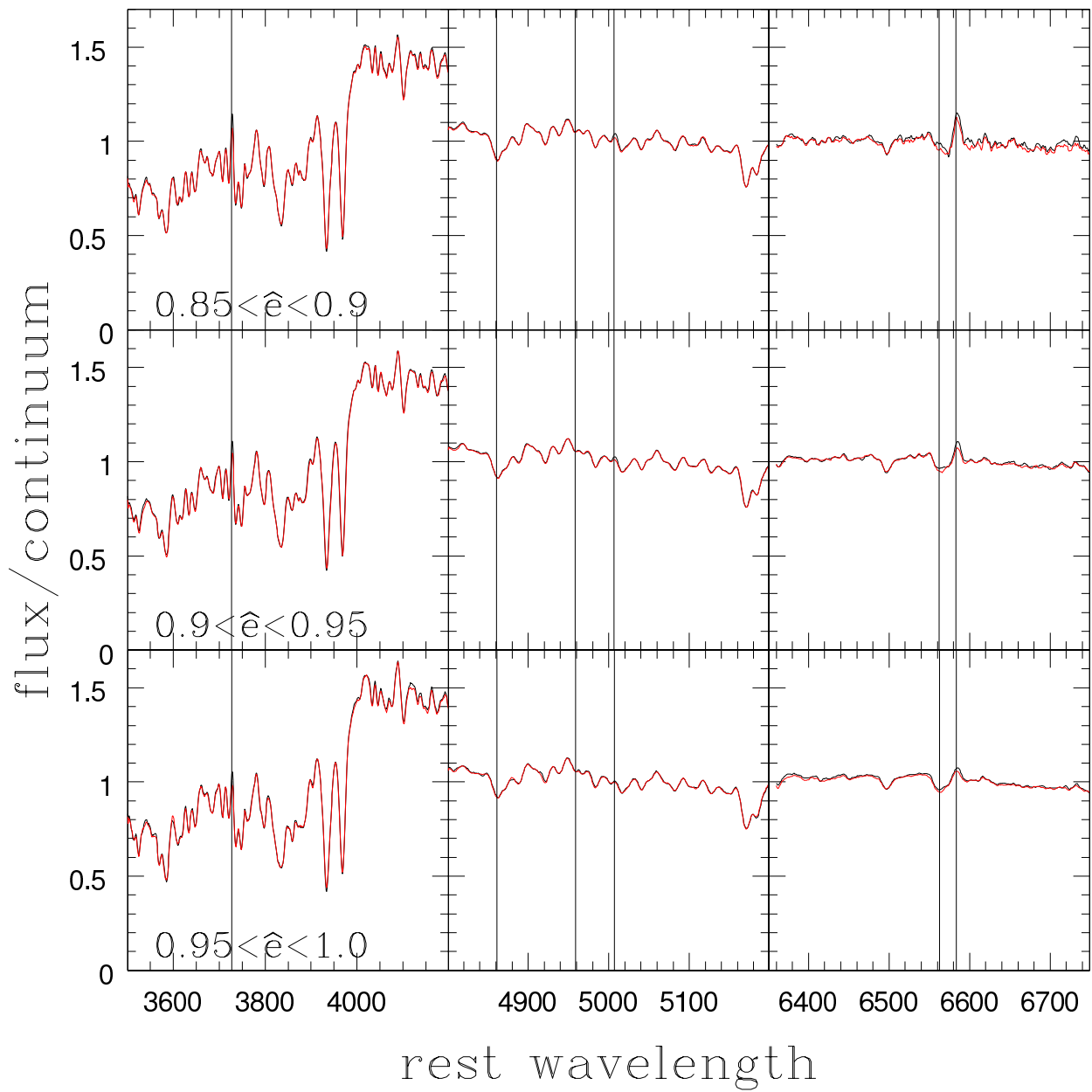}
\caption{Stacked spectra centered near the {\sc [oii]} (3727\AA; {\it left}),
{\sc [oiii]} (5007\AA; {\it middle}) and H$\alpha$/{\sc [nii]} 
(6563\AA/6583\AA; {\it right}) emission
lines.  Both the mean ({\it gray}) and median ({\it red}) spectra are
shown.  The constituent galaxies were taken from the $\hat{e}>0.85$
sample and divided into $\hat{e}$ bins as shown.  The spectra have been
normalized so the average value across the displayed range is 1.  Vertical
lines indicate important emission lines.  
\label{figSpectra}}
\end{figure}

In Figure \ref{figTemplates}, we show the average spectral energy densities
of the stacked $\hat{e}>0.925$ galaxies as a function of 
redshift, including the average optical, near-IR and mid-IR 
fluxes of the sources.  It is apparent that the
templates provide adequate fits in all redshift bins out to
$z=0.4$, but the last two redshift bins show discrepancies.  The
fit to the UV fluxes 
in the $z=0.6$ bin is particularly bad because increasing the
late-type contribution to fit the UV fluxes would overpredict the MIR
fluxes, which have smaller uncertainties.
This may be due, in part, to evolution in the shape of the UV excess
with look-back time (e.g. \citealt{brow00}).  Early-type
galaxies are also known to grow bluer and brighter with increasing
redshift due to their younger stellar populations 
(e.g. \citealt{ferr05}), which the
templates model by changing the relative contributions 
of the various components.  Changes in the intrinsic shape
of the stellar SEDs may lead to under-predicting the UV flux of
a passively-evolving stellar population.  We see some evidence for
this in Figure \ref{figEhatDist}, in which the $\hat{e}$-distribution
moves to lower $\hat{e}$ with redshift out to $z=0.4$, and in
Figure \ref{figTemplates}, which shows an excess of both UV and MIR
emission in the $z=0.6$ bin.

Figure \ref{figOptColor} shows the evolution of the 
measured $FUV-V$ colors (uncorrected for star formation) of 
our stacked galaxies.
The $FUV-V$ color of the uncorrected galaxies becomes moderately
bluer at higher redshifts,
in contrast with the results of R07, which suggest that 
color stays relatively constant with redshift.  The K-corrected colors,
shown in the lower panel, indicate that the rest-frame colors exhibit
no obvious evolution.

While the late-type templates never
contribute a significant fraction of the bolometric luminosity, averaging
only 7\%, their contribution to the UV flux can be significant. 
The colors shown in Figure \ref{figOptColor} do not account for 
contamination by star formation, so we must correct for the contribution
of star formation to the measured UV fluxes before attempting to
interpret the colors in the context of the UV excess.  If we assume that half
of the UV flux in all redshift bins beyond $z=0.2$ is contributed by star
formation, then a correction of 0.75 magnitudes to the $FUV-V$ colors
is required.  Such a correction would bring our colors 
into rough agreement with the colors reported in R07.  Since the
results of Figure \ref{figSpectra} indicate that $\hat{e}$ is
a reasonable tracer of contamination, we can use the
templates to estimate the contribution of young stars to
the measured UV flux.  These young stars may be due to
recent star formation or a component of the old stellar 
population that has not yet evolved off the main sequence.
At any given wavelength, the two late-type templates contribute a
fraction of the total flux, $r_{sf}$, given by
\begin{equation}\label{eqFluxRatio}
r_{sf}\left(\lambda\right) = \frac{f_{spiral}\left(\lambda\right)+f_{irregular}\left(\lambda\right)}{f_{ellip}\left(\lambda\right)+f_{spiral}\left(\lambda\right)+f_{irregular}\left(\lambda\right)}
\end{equation}
where $f_{x}\left(\lambda\right)$ is the density from template $x$ at 
wavelength $\lambda$.  If we assume that the UV 
flux from the elliptical
template has no contribution from young stars and that the UV fluxes
from the two star forming templates are contributed entirely by young stars,
we can then determine the $FUV-V$ colors
the stacked galaxies would have in the absence of young stars,
\begin{equation}\label{eqColorCorrection}
\left(FUV-V\right)_{corr} = \left(FUV-V\right)_{obs} - 2.5\log\biggl[1-r_{sf}\left(1550{\rm \AA}\right)\biggr]
\end{equation}
where $r_{sf}\left(1550{\rm \AA}\right)$ is the fraction of flux
contributed by star formation at the approximate center of the 
$FUV$ band.  The correction to the $V$-band magnitudes is negligible
for the relevant values of $\hat{e}$.
The corrections derived using this approach are not exact because the
star forming templates will likely have at least a small contribution
to their UV fluxes from hot, old stars and the elliptical template will
similarly contain a contribution from low-level star formation. 

\begin{figure}
\epsscale{1.0}
\plotone{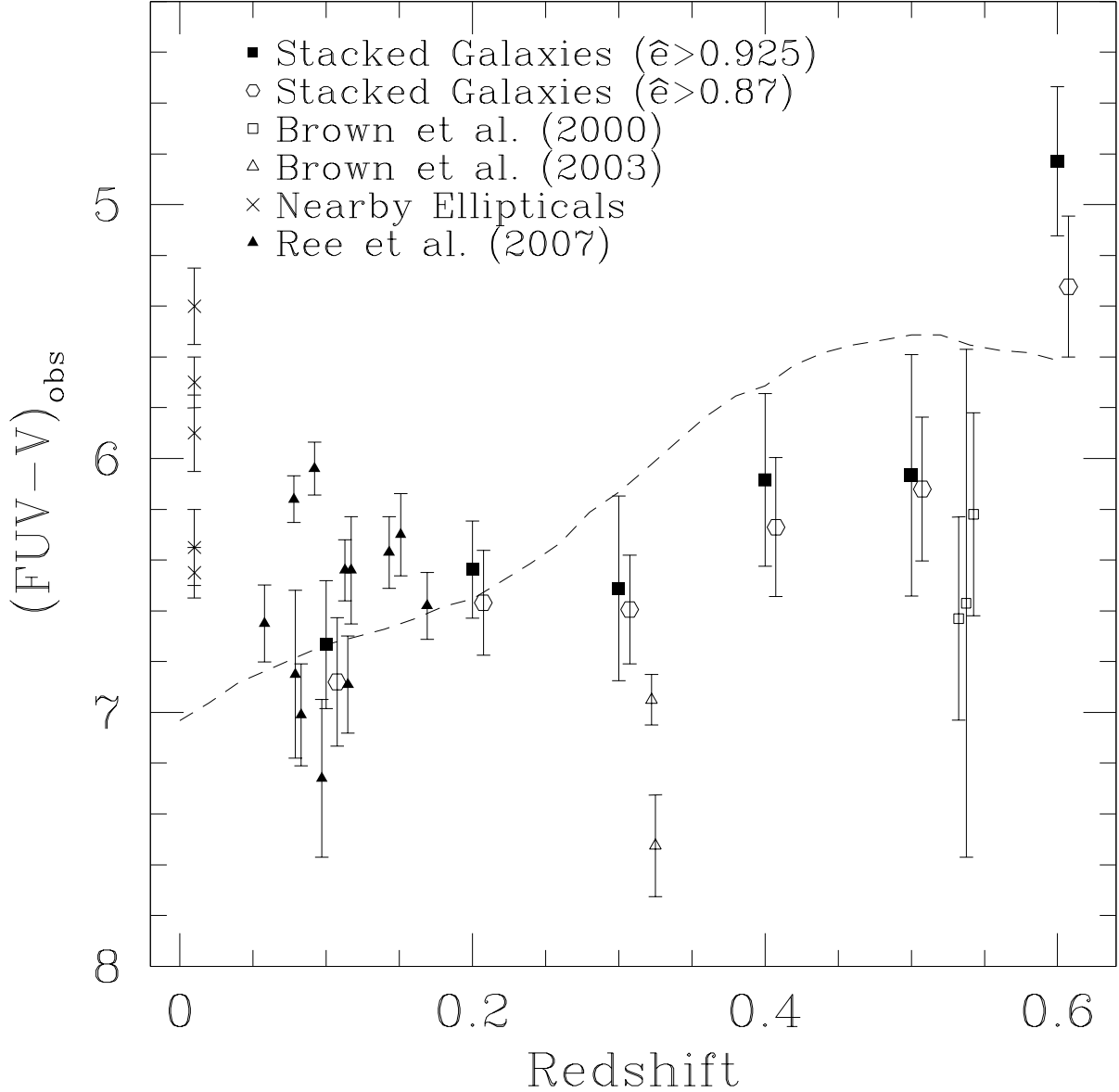}
\caption{Evolution of the observed colors of the
stacked elliptical galaxies, corrected for star formation.  
Flux contributed by young stars has been subtracted
using the method described in Eq. (\ref{eqColorCorrection}).  
The unevolving colors
({\it dashed line}) from Fig. \ref{figOptColor} have been translated
to pass through the $z=0.1$ point, but have not been otherwise
modified.  All other symbols
are the same as in Fig. \ref{figOptColor}.  Assuming Gaussian errors,
the $\hat{e}>0.87$ colors at $z\leq0.5$ are inconsistent with the 
non-evolving ``model'' at $>99\%$ confidence.
\label{figEvolutionCorrected}}
\end{figure}

If we use the corrected colors derived
from Eq. (\ref{eqColorCorrection}), our results agree
reasonably well with previous measurements, as shown in Figure 
\ref{figEvolutionCorrected}.  The resulting colors
from both the $\hat{e}>0.925$ and $\hat{e}>0.87$ samples
agree within the error bars, despite the different corrections for
star formation, so we conclude that our approach
is robust.  This agreement also suggests that our use of a fixed
$\hat{e}$ cut to select our galaxy sample does not significantly bias
our conclusions.
While the \citet{brow03} colors show a
slight disagreement with our results, the significant scatter about the
mean colors can likely account for the observed differences.  The
$z=0.6$ bin remains quite blue, probably because of the poor fit to the
UV fluxes (see Fig. \ref{figTemplates}).

Figure \ref{figEvolutionCorrected} indicates a systematic disagreement
between our high-z measurements and the non-evolving ``model'' for the
UV excess.  While the $\hat{e}$ data points are discrepant at only the
$1\sigma$ level, the considerably better statistics in the $\hat{e}>0.87$
sample lead to disagreement at the $2\sigma$ level in each redshift bin.
Disregarding the $z=0.6$ bin due to its poor fit, there is a 23\% (0.6\%)
probability of getting 3/5 data points to disagree with the model at
the $1\sigma$($2\sigma$) level.  If we also disregard the $z=0.5$ data point,
these probabilities increase to 38\% and 2\%, respectively.  This
indicates that, while our highest quality sample is consistent with an
unevolving model, the extended ($\hat{e}\geq0.87$) sample is not.  The
two samples return basically the same results, so we can reject an
unevolving model for the UV excess at $>$98\% confidence.

\begin{figure}
\epsscale{1.0}
\plotone{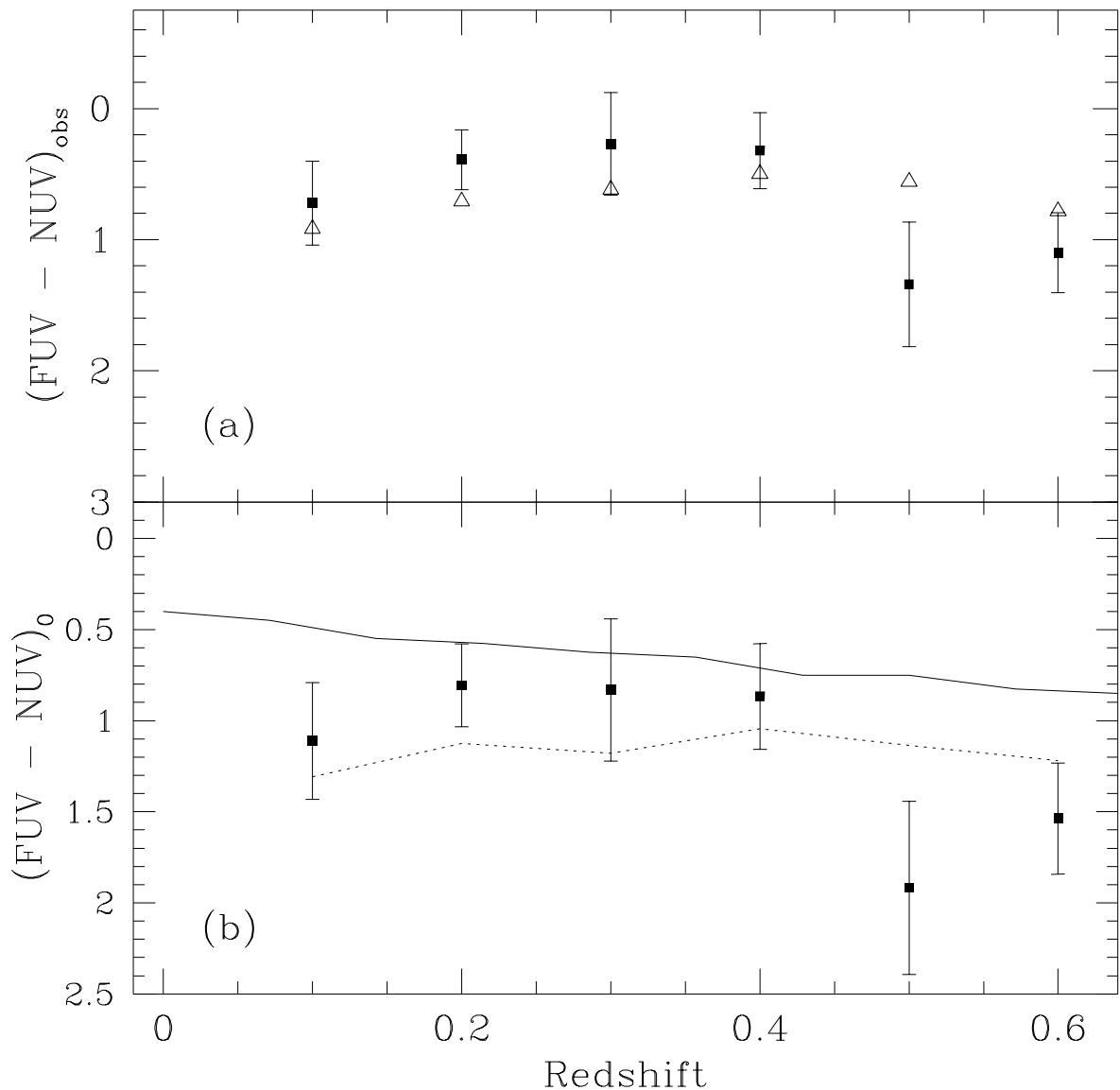}
\caption{Evolution of the UV color, uncorrected for star formation.  
Panel (a) shows the observer
frame evolution of UV color.  The {\it filled squares} give the 
UV colors of the stacked
galaxies, and the {\it open triangles} give the colors predicted by the
model spectra.  Panel (b) shows the rest frame evolution of UV
color, with error bars from the measured fluxes only.  
The {\it solid line} shows
the evolution predicted by the model of \citet{han07}, and the {\it dashed line}
shows the colors of the model galaxy spectra.  
\label{figUvEvolution}}
\end{figure}

In addition to considering the strength of the UV excess, we
examine its intrinsic shape by looking at the $FUV-NUV$ colors
of the stacked galaxies, as shown in Figure \ref{figUvEvolution}.  
Figure \ref{figUvEvolution}a shows that the
modified templates provide a reasonable approximation to the
shape of the UV excess over our entire redshift range.
The agreement between the measured and predicted colors in the $z=0.6$ bin
indicates that the disagreement between the measured fluxes and the model
spectrum in Figure \ref{figTemplates}
is one of normalization rather than shape.  We have not
corrected the colors in Figure \ref{figUvEvolution} for star 
formation because any such correction
will depend critically on the assumed shape of the UV excess, which we
are trying to measure.  Assuming that the early- and late-type templates give
a reasonable match to the UV excess and the emission from young
stars, respectively, the values of $r_{sf}\left(1550\right)$ 
and $r_{sf}\left(2250\right)$ should be similar and the correction small.  
(See Fig. \ref{figTemplates}.)
Figure \ref{figUvEvolution}b compares our K-corrected UV
colors to the prediction in HPL; the error bars in this panel do not
include the systematic uncertainties associated with applying K-corrections
from the model spectra.
The colors in Figure \ref{figUvEvolution}b appear to show
moderate evolution toward redder $FUV-NUV$ beyond $z=0.4$, 
and are inconsistent with the HPL
model.  However, if we use the differences between the measured $FUV-NUV$
colors and those predicted by the templates to estimate the 
systematic uncertainties associated with the templates, the error
bars on the $z=0.5$ and $z=0.6$ redshift bins increase by $\sim75\%$,
and we are no longer able to distinguish between the HPL model and an
unevolving spectral shape.

The principle difference between our work and existing studies is that
we examined a sample including all early-type galaxies above a fixed
luminosity cutoff rather than restricting the sample to galaxies
in rich clusters.  Since
galaxies in clusters are likely to be stripped of their gas, they are
also less likely to show recent star formation.  Different
star formation histories may also affect the metallicity distributions
of cluster ellipticals compared to those in the field, and either effect
could alter the UV properties of galaxies in rich clusters.

\begin{figure}
\epsscale{1.0}
\plotone{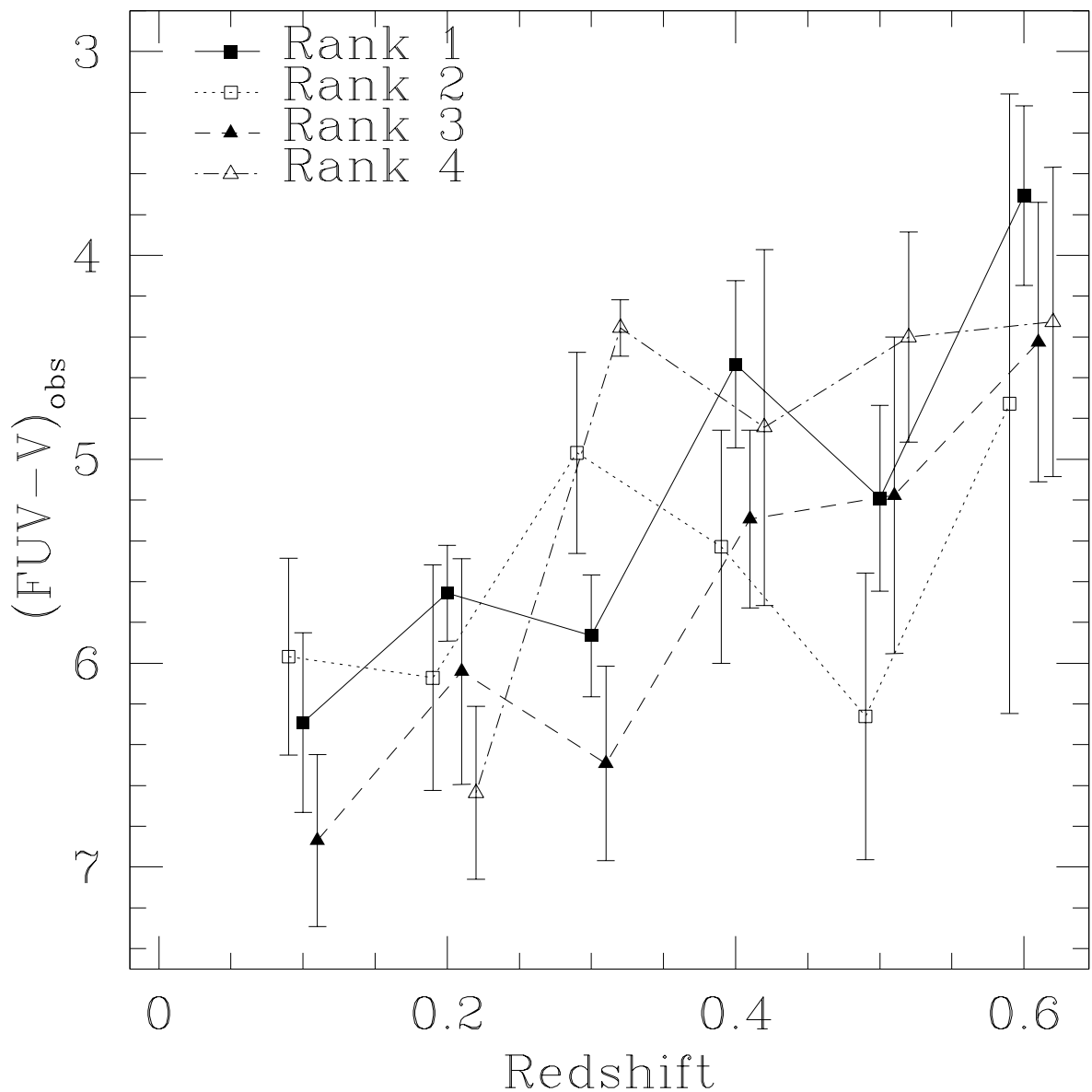}
\caption{Dependence of $FUV-V$ on galaxy environment.  Rank 1 galaxies
({\it filled squares}) fall in the least dense regions while Rank 4
galaxies ({\it open triangles}) fall in the densest regions.  The data show
no significant pattern, indicating that either a galaxy's environment has
little impact on its UV properties or
the influence of environment is limited to so few galaxies
that the other galaxies in the bin overwhelm any effect.
\label{figDensity}}
\end{figure}

We measure the effect of a galaxy's environment on its UV properties by
counting the number of bright elliptical galaxies within a projected
radius of $2{\rm h}^{-1}$ Mpc of each sample galaxy and 
within a photometric $\Delta z=0.03$, which is similar to the
resolution of our photometric redshifts.  We sorted the 
galaxies in our $\hat{e}>0.925$ sample in order of increasing
numbers of nearby bright ellipticals and divided the sample into four 
bins labeled 1-4 (low density to high density) with equal numbers 
of galaxies in each bin.  We stacked the galaxies in each density bin as 
described in \S\ref{secImage}, and the results are shown
in Figure \ref{figDensity}.  The displayed colors have not been
corrected for residual star formation, but since the different samples
have similar $\left<\hat{e}\right>$, the corrections
for the different density samples should also be similar.  
The trend toward bluer
$FUV-V$ color at higher redshift is found in all density bins,
although the scatter between the samples is sometimes significant.
We do not find any significant trend in $FUV-V$ as a function of environment.
This suggests that either the Horizontal Branch morphology does not
differ significantly between clusters and the field or the galaxies
that are significantly affected by their environment are so rare 
that the effects are overwhelmed by averaging with a 
large number of ``normal'' galaxies. 

\section{Summary and Conclusions}\label{secConclusion}
We have measured the evolution of the UV emission from the average
luminous, early-type galaxy with redshift by performing a stacking analysis 
with photometrically selected galaxies from the Bo\"otes field.
To the extent possible with the available data, we have eliminated AGN
from our sample, although the stacked AGES spectra suggest that weak
LINERs are ubiquitous.  LINERs show very weak UV continua compared to
broad-line AGN of similar luminosity, so the contamination of our
results due to nuclear activity is minimal.
We find that the observed $FUV-V$ colors of our stacked galaxies are
bluer than the colors of the BCGs studied by \citet{ree07},
\citet{brow00} and \citet{brow03} 
and show a pronounced tendency to become bluer with redshift.
The presence of a small excess in
8$\mu m$ emission indicates the need to correct for residual
star formation.  The necessity of such a correction,
even among a sample of galaxies that has been carefully selected to have
as little star formation as possible, suggests that contributions from
star formation should always be considered when measuring the UV excess
photometrically.  This is consistent with the results of other authors
(e.g. \citealt{kavi06}).

After correcting for star formation, we find that the intrinsic strength
of the UV excess, as measured by the $FUV-V$ colors, is inconsistent
with a non-evolving model.  The measured evolution shows reasonable 
agreement with 
previous studies.  However, the $FUV-V$ colors of the averaged galaxies 
remain modestly bluer than the individual cluster galaxies that have
been studied previously.  
Our results agree with the R07 results in the first two redshift 
bins, but their preferred models are inconsistent with our higher 
redshift data, suggesting that one or more of the model parameters
needs to be adjusted.  At least some of the evolution may be
due to an increase in $\langle L_{bol}\rangle$ from $5\times10^{10}{\rm L_{\odot}}$ 
to $8\times10^{10} {\rm L_{\odot}}$ as we go from $z=0.1$ to $z=0.5$,
since more luminous galaxies tend to show redder $FUV-V$.
The relatively good agreement between our results and the individual 
galaxies measured by other authors indicates that
our stacking analysis is able to probe the evolution of the UV excess
as well as detailed studies using small galaxy samples.

We also measured the evolution of the intrinsic shape of the UV excess,
finding little evidence for evolution in the rest-frame $FUV-NUV$ colors of
early-type galaxies in the 6 Gyr since $z=0.6$.  
Our two highest-redshift data points are slightly
redder than the others, possibly due to systematic effects
in our K-corrections, but the differences are not significant.
The UV colors of our stacked galaxies are also consistent
with the slow evolution predicted by HPL, but the change in the 
intrinsic $FUV-V$ color with redshift is inconsistent with their
predictions for the evolution in the UV-optical color.  While HPL
do not show any predictions for evolution in $FUV-V$, they do predict
the change in $FUV-r$, which should be quite similar to $FUV-V$, to be
only $\sim0.1$ mag.  (See HPL, Figure 3.)

We found evidence that the UV excess might weaken in more luminous,
and therefore more massive, galaxies.  The evidence for this trend is
marginal at best, however, and this result should be explored further.
If true, it would contradict the suggestion that the UV excess is 
stronger in more metal-rich galaxies, since the most massive galaxies tend
also to have the highest metallicities.  

We also divided our galaxies into subsamples by density to explore the
impact of environment on the UV excess, and
found that the trend for bluer $FUV-V$ colors at higher redshift 
occurs in all environments.  There is no identifiable trend in color
with density.  The lack of such trends is surprising, as
effects such as ram pressure stripping, felt by galaxies 
in rich clusters, could lead to changes in the age, mean metallicity 
or residual star formation rates of the constituent galaxies.

While several open questions remain, including the possible
variation in the UV excess with stellar mass, 
it appears that the average early-type galaxy evolves in roughly the same
way as the individual cluster galaxies measured thus far.
The measured evolution, along with
the lack of significant trends in color with galaxy environment,
should provide interesting constraints for
future models of stellar evolution.

\acknowledgments
We wish to thank Thomas Brown and Henry Ferguson for providing us
with their published UV spectra for a sample of elliptical galaxies; while
not used in the final text, these spectra were extremely useful for earlier
versions of this work.  We
are grateful to Tim Heckman for answering occasional questions
concerning the GALEX satellite and for helpful suggestions for ways to 
exclude possible instrumental effects as sources of systematic uncertainty.
Thanks are due Marc Pinsonneault, Michael Brown, Daniel Stern, Anthony
Gonzalez and an anonymous referee for their helpful comments on 
earlier drafts of this paper.  We
thank Rick Pogge for a useful discussion concerning the
our stacked spectra.
Additionally, we wish to thank the developers of the Funtools package of FITS
utilities, which we used extensively in our masking and
stacking routines.  We acknowledge the GALEX collaboration
for providing access to the DIS images used in this work.
This work made use of data products provided by the NOAO Deep Wide-Field Survey
(Jannuzi and Dey 1999; Jannuzi et al. 2005; Dey et al. 2005), which is 
supported by the National Optical Astronomy Observatory (NOAO).  NOAO is 
operated by AURA, Inc., under a cooperative agreement with the National 
Science Foundation.

\clearpage
\begin{deluxetable}{ccccc}
\tabletypesize{\scriptsize}
\tablewidth{0pc}
\tablecaption{GALEX DIS Pointings Overlapping the Bo\"otes Field
\label{tblPointings}}
\tablehead{
\colhead{Field Name} & \colhead{$\alpha_{J2000}$} & \colhead{$\delta_{J2000}$}&
\colhead{FUV Exposure Time(s)} & \colhead{NUV Exposure Time(s)}}
\startdata
NGPDWS\_00 & 219.15544 & +35.16978 & 89492 & 89492 \\
NGPDWS\_01 & 217.85994 & +35.41043 & 9508 & 9508 \\
NGPDWS\_02 & 219.20967 & +34.09978 & 5918 & 5918\\
NGPDWS\_03 & 218.15230 & +34.60758 & --- & 8798 \\
NGPDWS\_04 & 216.54480 & +35.45390 & --- & 8473 \\
NGPDWS\_05 & 217.20325 & +34.60071 & --- & 5493 \\
NGPDWS\_06 & 217.29972 & +33.64349 & --- & 9667 \\
NGPDWS\_07 & 216.12801 & +33.30057 & --- & 9650 \\
NGPDWS\_08 & 216.56646 & +32.26761 & --- & 8841 \\
NGPDWS\_09 & 218.92129 & +33.18095 & --- & 5801 \\
NGPDWS\_10 & 220.23785 & +34.75341 & 1107 & 1107 \\
NGPDWS\_11 & 220.40109 & +33.71991 & 1606 & 1606 \\
NGPDWS\_12 & 218.29527 & +33.92749 & --- & 8346 \\
NGPDWS\_13 & 217.85889 & +33.01259 & --- & 5343 \\
NGPDWS\_14 & 219.41794 & +32.39430 & --- & 7703 \\
NGPDWS\_15 & 216.39855 & +34.34794 & --- & 8783 \\
\enddata
\tablecomments{These exposure times are for the GR3 images.  With the
release of GR4, many of these fields now have deeper images available.}
\end{deluxetable}

\clearpage
\begin{deluxetable}{ccccccccc}
\tabletypesize{\scriptsize}
\tablewidth{0pc}
\tablecaption{Systematic Test Results
\label{tabSystematics}}
\tablehead{
\colhead{ } & \multicolumn{4}{c}{$FUV$} & \multicolumn{4}{c}{$NUV$} \\
\colhead{$z$} & \colhead{Mean} & \colhead{RMS} & 
\colhead{Bias} & \colhead{Bias RMS} & \colhead{Mean} & 
\colhead{RMS} & \colhead{Bias} & \colhead{Bias RMS}}
\startdata
0.1 & 22.97 & 0.42 & $-0.06$ & 0.05 & 22.18 & 0.26 & $-0.05$ & 0.06 \\
0.2 & 23.63 & 0.11 & $-0.01$ & 0.04 & 23.32 & 0.08 & 0.00 & 0.03 \\
0.3 & 24.53 & 0.06 & $-0.05$ & 0.01 & 24.10 & 0.06 & 0.05  & 0.03 \\
0.4 & 24.95 & 0.04 & 0.00 & 0.02 & 24.34 & 0.04 & 0.09  & 0.04 \\
0.5 & 25.19 & 0.03 & $-0.01$ & 0.01 & 24.37 & 0.04 & 0.15  & 0.02 \\
0.6 & 25.56 & 0.01 & 0.05  & 0.02 & 24.59 & 0.03 & 0.08  & 0.02 \\
\enddata
\tablecomments{The values of the RMS, Bias and Bias RMS listed here
were generated using 250 bootstrapped sampling
realizations.  The bias is defined as 
$\Delta m=\langle~m_{measured}-m_{predicted}\rangle$.  Here RMS indicates the
RMS scatter in bootstrapped sample mean.}
\end{deluxetable}

\clearpage
\begin{deluxetable}{ccccccccccccccc}
\tabletypesize{\scriptsize}
\tablewidth{0pc}
\rotate
\tablecaption{Extinction-Corrected Magnitudes
\label{tblPhot}}
\tablehead{
\colhead{} & \multicolumn{2}{c}{${\rm m}_{FUV}$} & \multicolumn{2}{c}{$\sigma_{FUV}$} &
\multicolumn{2}{c}{$FUV$ K-corr} & \multicolumn{2}{c}{${\rm m}_{NUV}$} & 
\multicolumn{2}{c}{$\sigma_{NUV}$} & \multicolumn{2}{c}{$NUV$ K-corr} & \multicolumn{2}{c}{$V$} \\
\colhead{$z$} & \colhead{$\hat{e}\geq0.87$} & \colhead{$\hat{e}\geq0.925$}
& \colhead{$\hat{e}\geq0.87$} & \colhead{$\hat{e}\geq0.925$} & \colhead{$\hat{e}\geq0.87$} & \colhead{$\hat{e}\geq0.925$}
 & \colhead{$\hat{e}\geq0.87$} & \colhead{$\hat{e}\geq0.925$} & \colhead{$\hat{e}\geq0.87$} & \colhead{$\hat{e}\geq0.925$}
 & \colhead{$\hat{e}\geq0.87$} & \colhead{$\hat{e}\geq0.925$} & \colhead{$\hat{e}\geq0.87$} & \colhead{$\hat{e}\geq0.925$}}
\startdata
0.1 & 23.55 & 23.46 & 0.25 & 0.25 & ${\rm -0.03}$ & ${\rm -0.03}$ & 22.74 & 22.74 & 0.19 & 0.20 & 0.42 & 0.36 & 17.12 & 17.12 \\
0.2 & 24.43 & 24.21 & 0.19 & 0.18 & ${\rm -0.05}$ & ${\rm -0.01}$ & 23.89 & 23.82 & 0.12 & 0.14 & 0.61 & 0.41 & 18.59 & 18.51 \\
0.3 & 25.52 & 25.32 & 0.17 & 0.30 & ${\rm -0.07}$ & ${\rm -0.03}$ & 24.92 & 25.05 & 0.10 & 0.25 & 0.73 & 0.53 & 19.77 & 19.47 \\
0.4 & 26.00 & 25.74 & 0.12 & 0.20 & ${\rm -0.12}$ & ${\rm -0.02}$ & 25.27 & 25.42 & 0.13 & 0.21 & 0.89 & 0.53 & 20.73 & 20.68 \\
0.5 & 27.00 & 27.30 & 0.20 & 0.41 & ${\rm -0.07}$ & 0.10 & 25.60 & 25.96 & 0.14 & 0.24 & 1.03 & 0.67 & 21.85 & 21.86 \\
0.6 & 27.03 & 26.69 & 0.23 & 0.23 & 0.12 & 0.32 & 25.60 & 25.59 & 0.18 & 0.20 & 1.05 & 0.76 & 22.43 & 22.31 \\
\enddata
\tablecomments{The uncertainties in $FUV$ and $NUV$ magnitudes are determined
using the dispersion about the mean of the bootstrapped magnitudes.  
K-corrections are computed using the routines of \citet{asse07} and the
modified spectral templates of \citet{asse08}.}
\end{deluxetable}

\clearpage

\clearpage

\clearpage

\clearpage

\clearpage

\clearpage

\clearpage

\clearpage

\clearpage

\clearpage

\end{document}